\documentstyle[floats,preprint,aps]{revtex} 
\input{epsf}
  
\begin{document} 
\draft
\title{A Thermodynamic Model for Receptor Clustering} 
 
\author{Chinlin Guo and Herbert Levine\footnote{Author to whom
correspondence should be addressed. Address: Department of Physics,
University of California, San Diego,  9500 Gilman Drive, La Jolla, CA
92093-0319.  Email: levine@herbie.ucsd.edu. Phone: (619)-534-4844.
Fax: (619)-534-7697.}}  \address{Department of Physics \\  University
of California, San Diego,  9500 Gilman Drive\\ La Jolla, CA
92093-0319\\}  \date{\today} \maketitle
\begin{abstract} 
Intracellular signaling often arises from ligand-induced
oligomerization of cell surface receptors. This oligomerization or
clustering  process is fundamentally a cooperative behavior between
near-neighbor   receptor molecules; the properties of this
cooperative process clearly affects the signal transduction.  Recent
investigations have revealed the molecular basis  of receptor-receptor
interactions,  but a simple theoretical framework  for using this data
to predict cluster formation has been lacking. Here, we propose   a
simple, coarse-grained, phenomenological model for ligand-modulated
receptor interactions and discuss its equilibrium properties via
mean-field theory. The existence of a first-order transition for this
model has immediate implications regarding the robustness of the
cellular signaling response.

\end{abstract} 

\pacs{Keywords: receptor clustering, phase diagram, thermodynamics}

\section{Introduction} 

Cell growth, differentiation, migration, and apoptosis are in part
regulated by extracellular polypeptide growth factors or cytokines
(Heldin, 1995; Stuart and Jones, 1995).  As these molecules are unable
to pass through the hydrophobic cell membrane,  they have to bind to
the extracellular domains of specific surface  receptors in order to
exert their effects. Much effort has gone into investigating the
fundamental question of how the ligand-receptor interaction can
trigger the proper intracellular signals.  One popular hypothesis is
that ligand-induced "clustering" of  ligand-receptor complexes can be
a key element in the proper activation of downstream signals.
(Ashkenazi and Dixit, 1998; Bray and Levin, 1998; Heldin, 1995;
Germain, 1997;  Lemmon and Schlessinger, 1994, 1998; Reich {\it et
al.}, 1997;  Sakihama {\it et al.}, 1995). 

As an example of this line of reasoning, we consider the signaling
cascade  mediated by the binding of tumor necrosis factor (TNF) to
the receptor TNF-R1. Internally, the cytoplasmic domain of TNF-R1 is
``sensed" by a variety of adaptor proteins, namely TRADD, FADD, TRAF2,
and RIP; this sensing leads eventually to NF$-\kappa$B/JNK/SAPK
activation and  apoptosis. To accomplish the downstream  signaling, an
oligomerization of these  adaptor proteins is required (Ashkenazi and
Dixit, 1998). One way to facilitate oligomerization is via
construction of a molecular scaffolding via TNF-induced TNF-R1
clustering. It is known that TNF-R1 will not aggregate in the absence
of TNF; this is due to the association of  an inhibitor, "silencer of
death domain"  (SODD), which normally attaches to TNF-R1 cytoplasmic
domains and prevents receptor aggregation (Jiang {\it et al.}, 1999),
or alternatively due to the receptor extracellular domains since
spontaneous association of TNF-R1 has been observed in cells  that
express truncated receptors (Boldin {\it et al.}, 1995; Vandevoorde
{\it et al.}, 1997). TNF treatment, however, can bring two or more
receptors into  proximity via its multiple binding capacity (Jones
{\it et al.},  1990, 1992)]. This ``proximity" might ``squeeze"  out
SODD (Jiang {\it et al.}, 1999), expose the  cytoplasmic ``death"
domains  to adaptor proteins, and thereby stabilize receptor clusters.
Thus, a molecular scaffold/nuclei is generated to initiate signaling.

Over a longer time scale, the signaling messages can provide feedback
to modify the capability of  surface receptor clustering  (Humphries,
1996; Wyszynski {\it et al.}, 1997). This leads to a complex dynamical
process involving both the intracellular signaling cascades as  well
as the  surface receptor clustering. The self-organization made
possible by these feedbacks has been intensively  discussed for
signaling cascades (see, e.g., Jafri and Keizer, 1995; Barkai and
Leibler, 1997). Much less is understood, however, regarding the role
of  receptor clustering. It is clear, though, that given the
hypothesis that  cellular signaling relies on the formation of
receptor  clusters, the temporal and spatial characteristics of
clustering  would certainly affect the process of signaling
transduction. Thus, modeling the physical properties of receptor
clustering is as important as  modeling signaling cascades.

Since clustering is due to an interaction between nearest-neighbor
receptors, it is obviously a cooperative process. From a physics
perspective a system with this type of cooperativity can exhibit a
first order  phase transition, corresponding to a jump in the surface
density  of ligand-receptor complexes. In the coexistence region of
this transition, the surface will spontaneously segregate into two
phases, dilute and dense.  This first order phase transition endows
the signal transduction process with the ability to produce a digital
signal in an analog world; this is independent  of the details of
intracellular cascades, instead, arising  from the intrinsic
cooperativity in ligand-receptor interaction. This has not been
adequately addressed in the few models studied to date  (Goldstein and
Wiegel, 1983; Goldstein and Perelson, 1984; Riley {\it et al.}, 1995;
Coutsias {\it et al.}, 1997; Shea {\it et al.}, 1997). 

The purpose of this work is to introduce a phenomenological model for
the TNF-TNFR1 system to describe the onset of receptor clustering
(phase separation).  Specifically, we assume that clustering can be
described by the  statistical mechanics of a simple lattice
Hamiltonian, incorporating the fundamental mechanism of a multimeric
binding capacity for the ligand.  We will calculate (via mean-field
theory) a phase diagram and show that clustering will be
thermodynamically favored for some range of  ligand and receptor
densities. Finally, we will do a simple Monte Carlo simulation of this
system, showing that receptor diffusion will lead rapidly to cluster
formation in the relevant parameter range.  We neglect the possibility
that there exist long-time feedbacks to modify the clustering
capacity, and we ignore some inessential details of the
receptor-ligand interaction.  More detailed models including these
effects, as well as applications to other signaling systems, will be
presented in the future.

\section{The Lattice Hamiltonian}

In our model, we treat the cell surface as a lattice with a spacing of
the order of a few $nm$; this is the closest that neighboring
receptors can get to each other. Each lattice site $i$ has either one
or zero receptor molecules, denoted as  $n_i=$ 1 or 0. Our receptor
has only two states: liganded or unliganded, and the interaction
between receptor molecules is determined by their states. This
"two-state" model is  over-simplified, yet we will see that it gives
reasonable predictions for the phase diagram. A ``state" label, $t_i=$
1 or 2, to represent unliganded or  liganded, then, can be assigned to
each occupied receptor. We will further assume that the only ligands
on the surface are those bound to receptors.  If we let the chemical
potential of the ligand be $\mu_L$ and that of the receptor be
$\mu_R$, we then get a contribution to the effective Hamiltonian of
the system 
\begin{equation}\label{hamiltonian1} 
H_1(\{n,t\})=-\sum_i\mu(t_i)n_i
\end{equation}
where $\mu(1)=\mu_R$ and $\mu(2)=\mu_R+\mu_L+g_L$ and $-g_L$ is the
binding energy between ligand and receptor.

We should clarify the relationship between the parameters used here
and those  in real experiments. Using standard ideas (Changeux  {\it
et al.}, 1967), we notice that with only this term, the partition
function can be factorized and reduced to a single site problem,
\begin{equation}\label{partition2}
Z=\prod_i Z_i=\prod_i \left[1+\sum_{k=1}^{2}e^{\beta\mu(k)}\right]
\end{equation}
From this, we can immediately obtain the expectation values of the
TNF-R1 concentration in the liganded and unliganded states. These  are
assumed to correspond to the equilibrium condition of the following
reaction (Corti {\it et al.}, 1994;  Grell {\it et al.}, 1998):
$\mbox{TNF-R1$^{\rm(m)}$+TNF}\rightleftharpoons
\mbox{TNF}\bullet\mbox{TNF-R1$^{\rm(m)}$}$, with a corresponding
equilibrium dissociation constant: $\frac{[{\rm
TNF-R1}^{\rm(m)}]_{eq}[{\rm TNF}]} {[{\rm TNF}\bullet{\rm
TNF-R1}^{\rm(m)}]_{eq}}=K_d^{tnf}\approx 0.59$ nM, where the notation
"TNF-R1$^{\rm(m)}$" means TNF-R1 molecule distributed on  the
artificial membrane, and where the brackets [$\dots$]$_{eq}$ indicates
the equilibrium concentration of the respective molecule.  From this,
we have $e^{\beta(\mu_L+g_L)}=\frac{\rm [TNF]}{K_d^{tnf}}$.  To obtain
the parameters individually, we might employ an "ideal gas law" for
the ligand.  This yields  $e^{\beta\mu_L}$=[TNF]$\left({h^2\over2\pi
m_{tnf}k_BT}\right)^{3/2}$, and $g_L=k_BT\ln\left[({2\pi
m_{tnf}k_BT\over h^2})^{3/2}/K_d^{tnf}\right]\approx 60k_BT$, with $h$
as Planck constant and $m_{tnf}$ as the mass of TNF.

We next add a receptor-receptor interaction term. This takes the
general form 
\begin{equation}\label{hamiltonian2} 
H_2(\{n,t\})=-{1\over2}\sum_{ij}J_{ij}a(t_i,t_j)n_in_j
\end{equation}
Here, $J_{ij}=1$ only when $i,j$ are nearest-neighbors and is
otherwise 0 (fig.\ref{sublattice}).  The function $a(t_i,t_j)$
indicates a ``state"-dependent interaction energy between
nearest-neighbor receptors, namely, $a(1,1)$ is the energy between two
unliganded receptors, $a(1,2)=a(2,1)$ is the energy between one
liganded and one unliganded  receptors, and $a(2,2)$ is the energy
between two liganded receptors.  We note that in general, higher order
terms might exist, especially considering the ``trimeric" nature of
the TNF ligand in our model problem.  We have similarly neglected the
details of the interactions of the ctyoplasmic domains, as per our
earlier discussion. Our goal is to elucidate the basic idea regarding
clustering in the simplest possible model, assured that adding more
details will not change the basic notion that there exists a
first-order transition due to the cooperativity.

$a(1,2)$ and $a(2,1)$ are the interaction energies, for which we will use
an effective binding
strength $g_E$ of the order of $g_L /10$, arising via one or two
hydrogen bonds between receptors. It is important to realize that
our simplified model does not treat explicitly the formation of multimers
via the multimeric binding. Instead, it arbitrarily assigns
the one ligand (binding two receptors into a dimer, e.g.) to one of
the receptors and describes the dimeric binding as an attraction
between a bound and an unbound receptor. Because of this, the model
cannot distinguish between this relatively strong interaction and the 
subsequent much weaker interaction between the dimers. In future
work, we will show that this complication does not alter the basic picture
presented here.

As discussed above, in the TNF system  there is
probably  a short-range and non-specific "excluding"  interaction
between two unliganded or two liganded (with different ligand
molecules) receptors.  For the sake of
simplicity, we will assume that the repulsive energy is of the same
order of the associative one, i.e.,  $a(1,1)\approx a(2,2)\approx
-g_E$. This assumption is not necessary, yet  it greatly simplifies
the mathematical task for analysis. 

The symmetry of $a(t_i,t_j)$ allows us to introduce a simple matrix
notation for the total Hamiltonian $H_1+H_2$.  If we use two-component
vectors for the state-labeling:
$\tau_i=\left[\begin{array}{c}1\\0\end{array}\right]$ for $t_i=1$, and
$\tau_i=\left[\begin{array}{c}0\\1\end{array}\right]$ for $t_i=2$,
then the  Hamiltonian can be rewritten as
\begin{equation}\label{hamiltonian2a}
H(\{n_i,\tau_i\})=-\sum_in_i[\mu(1),\;\mu(2)]\tau_i -{1\over2}g_E
\sum_{ij}J_{ij}n_in_j\tau_i^+\left[\begin{array}{cc}-1&1\\1&-1\end{array}
\right]\tau_j
\end{equation}
Here $[\mu(1),\;\mu(2)]$ is an $1\times2$ matrix.  The simplicity of
using this form of the matrix $a(t_i,t_j)$ can immediately be seen if
we make a transformation  
 $\tau_i ={1\over2}\left[ \begin{array}{c}
 1+\sigma_i\\ 1-\sigma_i\end{array}\right]$, 
with $\sigma_i =\pm1$. Then
\begin{equation}\label{Hamiltonian3}
H(\{n_i,\sigma_i\})=-x\sum_in_i-y\sum_in_i\sigma_i
+{1\over2}g_E\sum_{ij}J_{ij}n_in_j\sigma_i\sigma_j
\end{equation} 
where $x=[\mu(1)+\mu(2)]/2$ is the "averaged" receptor chemical
potential, and  $y=[\mu(1)-\mu(2)]/2$ is directly related to the
ligand concentration, $e^{-\beta y} =\sqrt{{[{\rm TNF}]\over K_d^{tnf}}}$. 
The partition function then reads
\begin{equation}\label{partition0}
Z=\sum_{\{n_i,\sigma_i\}}\exp[-\beta H(\{n,\sigma\})]
\end{equation}
where $\sum_{\{n_i;\sigma_i\}}$ means ensemble summation over the
three  different configurations $\{n_i=0; n_i=1,\sigma_i=\pm1\}$ on
each lattice site, $\beta=1/k_BT$ with $k_B$ as Boltzmann factor and
$T$ as temperature.

If we define a new notation $u_i=n_i\sigma_i$, our model
would be very similar to a  spin-1 antiferromagnetic (AFM) BEG model
(Blume {\it et al.}, 1971), 
\begin{equation}\label{Hamiltonian4}
H(\{u_i\})=-x\sum_iu_i^2-y\sum_iu_i+{1\over2}g_E\sum_{ij}J_{ij}u_iu_j
\end{equation}
The origin of this AFM behavior is the ``negative cooperation'' between
nearest-neighbor receptors, as we have imposed that a ``proximity" of
two unliganded or two liganded receptors will cost energy.  Similar
behavior might occur in the erythropoietin receptor (EPO-R) and the
human Growth hormone receptor (hGH-R) systems (Heldin, 1995). This
negative cooperation will give rise to an  absence of clustering in
extreme high/low ligand concentration (i.e., $y\rightarrow\pm\infty$),
and thereby result in a ``bell''-shape or window-like 
signaling response (Elliott {\it et al.}, 1996). 

We should point out that this negative cooperation is not 
universal. In the case of EGF-R (epidermal growth factor receptor) system, a 
ferromagnetic (FM) behavior (``positive cooperation'') is more likely, since 
there clustering requires two or more liganded receptors (Lemmon {\it et al.}, 
1997). Thus the higher the ligand concentration, the more the EGF-R cluster 
can be formed, and  the EGF-EGFR signaling response behaves in a 
sigmoidal rather than a window-like pattern. It is clear that in both EGF-R and 
TNF-R (and hGH-R, EPO-R) systems, the ligand multiple binding capacity is the 
essential ingredient to induce clustering (of course one should consider the 
effect from receptor cytoplasmic domain as well). Which  kind of 
cooperation (negative or positive) one should one consider depends 
on the details of the receptor-receptor interaction (also including the 
chemical modifications on 
receptor cytoplasmic domains), and needs to be be established experimentally.
But, the essential feature of a first-order-transition-like behavior
in receptor clustering is not dependent on the sign of this 
additional cooperativity.
 
\section{Numerical Simulation}

To see if our model can generate clustering, we perform a Monte Carlo
simulation on a square lattice with the standard Metropolis scheme. For
simplicity, we fix the number liganded and unliganded receptors and
do not allowed these to fluctuate.  Given the rather strong binding, this is not an important
constraint. Furthermore, we allow motion only for individual recpetors
and do not explicitly allow a clusters to move as a whole; this
might not be the case in reality. 
The ``jumping" probability for a receptor to move to another lattice
site is determined by the Hamiltonian and obeys the detailed balance
law. In detail, we pick a receptor at random and try to move it
in a randomly chosen direction. The move is accepted if it lowers the
energy and the move is accepted with probability $e^{-\beta \Delta H}$
if the energy increases. 

From fig.\ref{monte}, we immediately see that for a given receptor
density, changing the ligand concentration  moves the system from a
non-clustering to a clustering phase. In this figure, the open, filled
circles indicate liganded, unliganded receptor molecules, respectively.
Note that the open and filled circles are arranged in an alternative way
to form the cluster (i.e., inside a cluster, the nearest neighbors of
the open circles must be filled circles, and vice versa). This implies
that the equilibrium state (which must be translationally invariant), 
can be described via dividing the system into 
two interleaved sub-lattice systems: one sub-lattice is occupied by one species 
of receptor molecule (liganded or unliganded), and all its nearest neighbors 
belong to the alternative sub-lattice which is occupied by another species. 

To obtain more insight into the conditions where receptor clustering can 
take place, we next analyze the partition function via the mean-field 
approximation.
 
\section{Mean field approximation} 

To proceed, we decouple the quadratic term in the Hamiltonian by
introducing an auxiliary Gaussian field and employing the standard
Hubbard-Stratonovich/Gaussian transformation (see, e.g., Amit, 1993;
Parisi, 1988), eqn.(\ref{Gaussian}). The benefit of this
transformation is to decouple the quadratic terms into linear terms
such that we can sum over the  ensemble configuration
($\{n_i,\sigma_i\}$) at each lattice site $i$  independently. This
yields (see Appendix for details)
\begin{eqnarray}\label{partition3}
Z&=&C\int {\cal D} \phi e^{{\beta
g_E\over2}\sum_{ij}J_{ij}\phi_i\phi_j+\sum_i S_i(\{\phi\})}\\
\mbox{with}\hspace{.2in} S_i(\{\phi\})
&=&\ln\left[1+2z\cosh(\beta[g_E\sum_jJ_{ij}\phi_j+y])\right]\nonumber
\end{eqnarray}
where ${\cal D}\phi=\prod_i d\phi_i$, and $C$ is a normalization
constant which does not affect the thermodynamic properties of the
partition function. The new field $\phi$ ranges from  $-\infty$ to
$+\infty$,  $y=-(\mu_L+g_L)/2$, and $z=e^{\beta
x}=e^{\beta[\mu_R+(\mu_L+g_L)/2]}$  is related to the receptor
chemical potential which remains to be determined (in terms of the
receptor density). The  first term in  eqn.(\ref{partition3}) is
related to the interaction energy between  nearest-neighbor lattice
points, whereas the second term is related  to the entropy arising due
to the available configurations on an individual lattice site.

In mean-field theory, we try to determine a ``homogeneous" saddle
point  approximation for the partition function.  For our system, the
negative  cooperation (i.e. the AFM nature) suggests that the system
might prefer having neighboring sites in oppositely liganded states.
Thus, we  separate the lattice into two interleaved sub-lattice
systems: all nearest-neighbors of a lattice site belong to the
alternate  sub-lattice (fig.\ref{sublattice}). We then assign two
``uniform" order parameters, $\phi_{\pm}$ to each sublattice.  After
this assumption, the exponent of the Boltzmann factor in the partition
function [eqn.(\ref{partition3})] now becomes $\rightarrow{\rm
{N\over2}}\left[\beta g_ED\phi_+\phi_- +S(\phi_+,\phi_-)\right]$,
where N is the number of total lattice  sites,
$S(\phi_+,\phi_-)=\sum_{k=\pm}\ln[1+2z\cosh(\beta[g_ED\phi_k+y])]$,
and $D$ is the number of nearest neighbors, which depends on the
structure of lattice. For instance, a square lattice yields $D=4$,
whereas a  honeycomb lattice yields $D=3$. 

Next, we minimize the free energy by varying $\phi_{\pm}$.   The
variation yields the ``saddle point" equation
$$\left\{{\partial\over\partial\phi_{\pm}}\Bigl[\beta g_ED\phi_+\phi_-
+S(\phi_+,\phi_-)\Bigr]\Bigl|_{\phi_{\pm}=\widetilde{\phi}_{\pm}}\right\}=0
$$. Working this out explicitly, we find  a self-consistent equation
for $\widetilde{\phi}_{\pm}$  
\begin{equation}\label{mf1}
\widetilde{\phi}_{\pm}=-\frac{2z\sinh[\beta(g_ED\phi_{\mp}+y)]}
{1+2z\cosh[\beta(g_ED\phi_{\mp}+y)]}
\end{equation}
with the free energy density 
\begin{equation}\label{f1}
f(\widetilde{\phi}_+, \widetilde{\phi}_-, z)
=-{g_ED\over2}\widetilde{\phi}_+\widetilde{\phi}_-
-{k_BT\over2}\sum_{k=\pm}\ln[1+2z\cosh(\beta[g_ED\widetilde{\phi}_k+y])]
\end{equation}
Finally, the mean-field receptor density is given by $\langle
n\rangle=-\partial f(\widetilde{\phi}_+,  \widetilde{\phi}_-,
z)/\partial \mu_R$. Explicitly, we have
\begin{equation}
\langle n\rangle \ = \ \sum_{k=\pm}  \frac{z
\cosh(\beta[g_ED\widetilde{\phi}_k+y])}  {1+2z
\cosh(\beta[g_ED\widetilde{\phi}_k+y])}
\end{equation}
We can therefore determine the receptor chemical  potential, $x$ (or
equivalently $z$) in terms of $\langle n\rangle$.  Thereafter, we can
rewrite the free energy density in  terms of $\langle n\rangle$,
$\widetilde{\phi}_+$, and $\widetilde{\phi}_-$.

\section{The onset of clustering}

There is no closed form solution for eqn.(\ref{mf1}). To get some
analytical information, we define
$\widetilde{\phi}_{\pm}=m\pm\epsilon$ and, with
$U(w)=\frac{2z\sinh[\beta(g_EDw+y)]}{1+2z\cosh[\beta(g_EDw+y)]}$,  we
have
\begin{equation}\label{mf2}
m=-\sum_{k=0}^{\infty}\frac{\epsilon^{2k}}{(2k)!}  U^{(2k)}(m)
\end{equation}
\begin{equation}\label{mf3}
\epsilon=\sum_{k=0}^{\infty}\frac{\epsilon^{2k+1}}{(2k+1)!}
U^{(2k+1)}(m)
\end{equation}
where $U^{(k)}(w)=d^kU(w)/dw^k$. The basic idea of separating out the
$\epsilon$ dependence is that solutions with non-zero values of
$\epsilon$ represent phases in which the proximity of neighboring
receptors gives rise to alternating ligand binding. For very small
receptor densities, there are few neighboring receptors and hence we
expect to find a unique solution of the mean-field equations with
$\epsilon=0$. In fact, it is clear from eqn. (\ref{mf3}) that there is
a solution with $\epsilon=0$ for all values of the parameters, but at
larger densities, there may be other more stable phases. The goal of
our analysis will be to understand the general structure of the phase
diagram and then to  obtain more quantitative detail by numerical
means.
 
To proceed, let us assume that $\epsilon$ is small and solve
eqns.(\ref{mf2}),  (\ref{mf3}) to order $\epsilon^2$;
\begin{eqnarray}\label{mf4}
m&=&m_0-m_1\epsilon^2 \\ \mbox{with}\;\;m_0&=&-U^{(0)}(m_0)
\nonumber\\ m_1&=&\frac{U^{(2)}(m_0)} {2!\left[1+U^{(1)}(m_0)\right]}
\end{eqnarray}
\begin{equation}\label{mf5}
\epsilon^3=\frac{\left\{U^{(1)}(m_0)-1\right\} \epsilon} {m_1
U^{(2)}(m_0)-\frac{1}{6} U^{(3)}(m_0)}
\end{equation}
Using the relationship given above for $\langle n\rangle$,  It is easy
to verify that $U^{(1)}(m_0)=\beta g_E D [\langle n\rangle -m_0^2]$,
$U^{(2)}(m_0)=-m_0 (\beta g_E D)^2 [1-3\langle n\rangle+2m_0^2]$, and
$$
U^{(3)}(m_0)= \left( \left[ 1-3\langle n\rangle+6m_0^2 \right]  \left[
\langle n\rangle-m_0^2 \right] -3m_0^2 \left[ 1-\langle n\rangle
\right] \right) \   \left( \beta g_E D \right)^3
$$ 

We must consider separately the cases where the denominator of
eqn. (\ref{mf5}) is positive or negative. Let us first imagine it is
positive, Then, the existence of a non-trivial solution of
eqn. (\ref{mf5}) requires that   $\left\{\beta g_ED[\langle n\rangle
-m_0^2]-1\right\}>0$.  At small $\langle n\rangle$ this condition will
clearly fail and we will have only the trivial solution.  Also, this
condition will fail at $\langle n\rangle$ close to 1 for large enough
$|y|$. We can see this by comparing the equation for $m_0$ with the
expression for $\langle n\rangle$. Note that if $y$ is large enough
such that the hyperbolic functions can be replaced by exponentials, we
have $|m_0| = \langle n\rangle$, and the above expression can be
replaced by  $\left\{\beta g_ED[\langle n\rangle -\langle n\rangle
^2]-1\right\}$; this is negative for the stated condition. As we cross
a line in parameter space such that this factor changes sign to
positive, there will be new  solutions at non-zero $\epsilon^2$ and
the one at $\epsilon=0$ becomes a local maximum of the free
energy. This emergence of a double-well structure, with a continuous
growth the non-zero $\epsilon^2$ solution indicates that the system
exhibits a second-order phase transition. 

We must next take into account the possibility that  $\left\{m_1 U^{(2
)}(m_0) - \frac{1}{6} U^{(3)}(m_0)\right\}<0$. Having the denominator
cross zero gives rise in our current approximation to a large value of
$\epsilon$ which thus invalidates the neglect of higher-order
terms. Typically, the higher-order terms will stabilize the system at
some finite value of $\epsilon$, which thus appears ``spontaneously"
as some parameter threshold is crossed. This is a first-order phase
transition, or equivalently a triple-well structure for the free
energy.  If the local minima (for zero and non-zero $\epsilon^2$) have
equally low free energy density, the system can exist in a mixture of
the two phases.  As we will see, the two coexisting phases differ in
their receptor density. Finally,  the points where both  $\left\{\beta
g_ED[\langle n\rangle -m_0^2-1]=0\right\}$ and $\left\{m_1 U^{(2
)}(m_0) -\frac{1}{6} U^{(3)}(m_0) \right\}=0$, are  ``critical
end-points" points, since they correspond to places where a
second-order transition line ends at a first-order line. A diagram of
this behavior, generated by the numerical solution of the mean-field
equations,  is given in fig. \ref{phase_diagram}.

For a given ligand concentration, we can find the phase coexistence
lines arising due to the first-order phase transition. This is done by
finding two solutions (solved with differing values of $\epsilon^2$)
of the mean field equations and then fixing $z$ (as a function of $y$)
by requiring that they have equal free-energy
\begin{equation}\label{binodal}
f(\phi^{(d)}, \phi^{(d)}, z)=f(\phi_-^{(c)}, \phi_+^{(c)}, z)
\end{equation}
where $\phi_{\pm}^{(c)}$, are the order parameters for the dense
condensed phase and $\phi^{(d)}$ the (equal) ones for the dilute
phase. For the condensed phase, the receptor density is close to unity
for reasonable values of the cooperativity parameter $\beta g_ED$. The
workings of this system as far as signaling is concerned is shown in
fig. {\ref{phase_separation}. Assume there is some fixed value of the
receptor density. As the ligand concentration is increased, we will
cross the phase transition boundary and the receptors will segregate
into a condensed phase and a dilute one, corresponding to the two
co-existing mean-field solutions. Under our basic hypothesis that
signaling is effected by having dense clusters, the response will
exhibit a sharp jump at a specific threshold ligand concentration.
Similarly, as the ligand concentration becomes too high we cross back
to the uniform receptor density state and signaling ceases. That is,
we have a ligand concentration ``window" for receptor clustering.

As can seen from the figure, the ``clustering" window will cease to
exist  below some minimal receptor density, as we never enter the
phase coexistence  region. By symmetry, this minimal density can be
found by solving the mean  field equations for $y=0$ where $m=0$. This
leads after some algebra to the self-consistent equations  
\begin{eqnarray}\label{n_min}
\langle n\rangle_{min}^{(d)}&=&\frac{e^{\beta g_ED\epsilon^2/2}-1}
{\cosh[\beta g_ED \epsilon]-1}\\ \mbox{with}\; \epsilon&=&\langle
n\rangle_{min}^{(d)} \tanh[\beta g_ED \epsilon] \nonumber
\end{eqnarray}
The numerical solution of these equations is presented in
fig.\ref{nmin}.  As the cooperativity parameter is increased, the
minimum density which will support a clustering window goes rapidly to
zero.  For the  TNF-TNFR1 cluster, it has been speculated that the
structure of cluster is a  honeycomb-like lattice (Bazzoni and
Beutler, 1995; Naismith {\it et al.}, 1995, 1996), which implies the
number of nearest-neighbor $D=3$. If we use our rough estimate $g_E
\simeq g_L/10 \simeq 6kT$, we find that $\langle
n\rangle_{min}^{(d)}\lesssim 10^{-6}/a_0^2$. Here $a_0$ is length scale of the
lattice spacing. If we take $a_0\approx 1$ nm, on a cell with surface area
100$\mu m^2$, this estimate yields  a requirement for less  than $10^2$
TNF-R1 molecule distributed on the cell surface. Given that an
average number of expressed TNF-R1 on cell surface is $\sim2000$, we
find that the cell operates within the desired  part of the phase
diagram and hence should exhibit strong sensitivity to the application
of TNF.  However, we should point out that this estimate is very
rough, as we have made a number of simplifying assumptions and this
issue needs to be re-visited with a more precise model of the receptor
interactions.

\section{Discussion} 

We have presented a simple model for signal transduction via receptor
clustering, based loosely on the TNF-TNFR1 system. Our basic idea is
simple. The interaction between receptors can lead to a first-order
phase transition with a discontinuous jump in the receptor density as
a function of the receptor chemical potential and/or the ligand
concentration.  Turning this around, this implies that the receptor
system will spontaneously phase separate for a range of ligand
concentrations. This fact about the thermodynamic equilibrium state
will lead under reasonable kinetic assumptions to the rapid formation
of receptor clusters. Assuming that these clusters are necessary for
the signal to proceed downstream has the immediate consequence that
the system exhibits a strong robust response independent of any
details of the  intracellular signaling cascade. This might provide a
simple solution to  the problem faced by biological evolution of how
to get digital response in an analog world.

From a physics perspective, there is nothing very surprising about our
phase diagram findings.  The idea of a "lattice" Hamiltonian  with
intrinsic "cooperativity"  has been proposed before (Changeux {\it et
al.}, 1967), and on general grounds models of this sort can be
expected to have  first order phase transitions. What is new here is
the connection of the transition to signaling via the idea of receptor
clustering. This connects nicely with increasing evidence that
clustering is ``universal" among many types of receptor classes.

In our model, we have ignored more-than-two receptor interaction,  and
relevant internal chemical degrees of freedom (such as the
dissociation of  SODD in the TNF-R1 system). We do not expect these
detailed considerations to change the overall  picture, but a more
sophisticated model will be needed to make more quantitative estimates
of ligand thresholds, cluster structures and most interestingly,
clustering dynamics.  We hope to report on these issues in the future,
as well as on the extension of our models to other ligand-receptor
systems.

Finally, it would be important to extend our work to later-stage
dynamics, as  that would allow the consideration of processes such as
adaptor protein-mediated  receptor internalization,
cytoskeleton-assisted cluster stabilization,  receptor affinity
regulation, receptor cross talk, and adaptation  (Barkai and Leibler,
1997; Hahn {\it et al.}, 1993;  Humphries {\it et al.}, 1996;
Holsinger {\it et al.}, 1998;  Luo and Lodish, 1997; Stewart {\it et
al.}, 1998; Sundberg and Rubin, 1996; Valitutti {\it et al.}, 1995;
Wyszynski {\it et al.}, 1997).  Other possible extensions might
involve the inclusion of spatial fluctuations, the explicit treatment
of external perturbations (Shoyab and Todaro, 1981),  the local
heterogeneity of the micro-environment (Bean {\it et al.}, 1988;  Ward
and Hammer, 1992), or fluctuations of ligand concentration; all of
these  issues have been neglected here.

\section{acknowledgment} 
Chinlin Guo wishes to acknowledge the LJIS Interdisciplinary Training
Program and the Burroughs Wellcome Fund for fellowship support. He
also acknowledges Margaret Cheung for the help with the numerical
simulation.  Herbert Levine acknowledges the support of the US NSF
under grant DMR98-5735.

\appendix
\section{The Gaussian Transformation}

The identity 
\begin{equation}\label{Gaussian0}
\int_{-\infty}^{\infty}dx \exp\left[-{1\over2g}x^2+isx\right]
=C\times \exp\left[-{g\over2}s^2\right]
\end{equation}
with $g>0$, $i=\sqrt{-1}$, and $C$ as some constant, can be generalized to 
\begin{equation}\label{Gaussian1}
\int_{-\infty}^{\infty}\prod_idx_i
\exp\left[-{g\over2}\sum_{jk}x_jJ_{jk}x_k+ig\sum_js_j\sum_kJ_{jk}x_k\right]
=C\times \exp\left[-{g\over2}\sum_{ij}s_iJ_{ij}s_j\right]
\end{equation}
as long as $J$ is a symmetric positive definite matrix. Thus, 
eqn.(\ref{partition3}) can be obtained by
\begin{eqnarray}\label{Gaussian}
Z&=&\sum_{\{n_i,\sigma_i\}}\exp\left\{\beta[x\sum_in_i+y\sum_in_i\sigma_i
-{g_E\over2}\sum_{ij}J_{ij}n_i\sigma_in_j\sigma_j]\right\} \\
&=&\sum_{\{n_i,\sigma_i\}} 
C_{\theta}
\int \prod_j \theta_j e^{-{\beta g_E\over2}\sum_{jk}J_{jk}\theta_j\theta_k
+\sum_j\beta[x+\sigma_j(y+ig_E\sum_kJ_{jk}\theta_k)]n_j}\nonumber\\
&=&C_{\phi}\int \prod_j\phi_j e^{{\beta g_E\over2}\sum_{jk}J_{jk}\phi_j\phi_k}
\times \left\{\sum_{\{n_j,\sigma_j\}} e^{ \sum_j
\beta[x+\sigma_j(y+g_E\sum_kJ_{jk}\phi_k)]n_j} \right\} \nonumber\\
&=&C_{\phi}\int \prod_j\phi_j e^{{\beta g_E\over2}\sum_{jk}J_{jk}\phi_j\phi_k}
\times \prod_j\left\{\sum_{n_j,\sigma_j} e^{ 
\beta[x+\sigma_j(y+g_E\sum_kJ_{jk}\phi_k)]n_j} \right\} \nonumber\\
&=&C_{\phi}\int \prod_j\phi_j e^{{\beta g_E\over2}\sum_{jk}J_{jk}\phi_j\phi_k}
\times\prod_j\left\{1+2z\cosh\Biggl[\beta \Bigl(g_E\sum_kJ_{jk}\phi_k+y
\Bigr)\Biggr]\right\}\nonumber\\
&=&C_{\phi}\int \prod_j\phi_j e^{{\beta g_E\over2}\sum_{jk}J_{jk}\phi_j\phi_k
+\sum_jS_j(\{\phi\})}\nonumber
\end{eqnarray}
where $i=\sqrt{-1}$, $\phi_k=i\theta_k$, 
$S_j(\{\phi\})=\ln \left[1+2z\cosh(\beta [g_E\sum_kJ_{jk}\phi_k+y])\right]$, 
and $C_{\theta}$, $C_{\phi}$ are integral constants.

\begin{figure}\centerline
{\epsfxsize = 6.in \epsffile{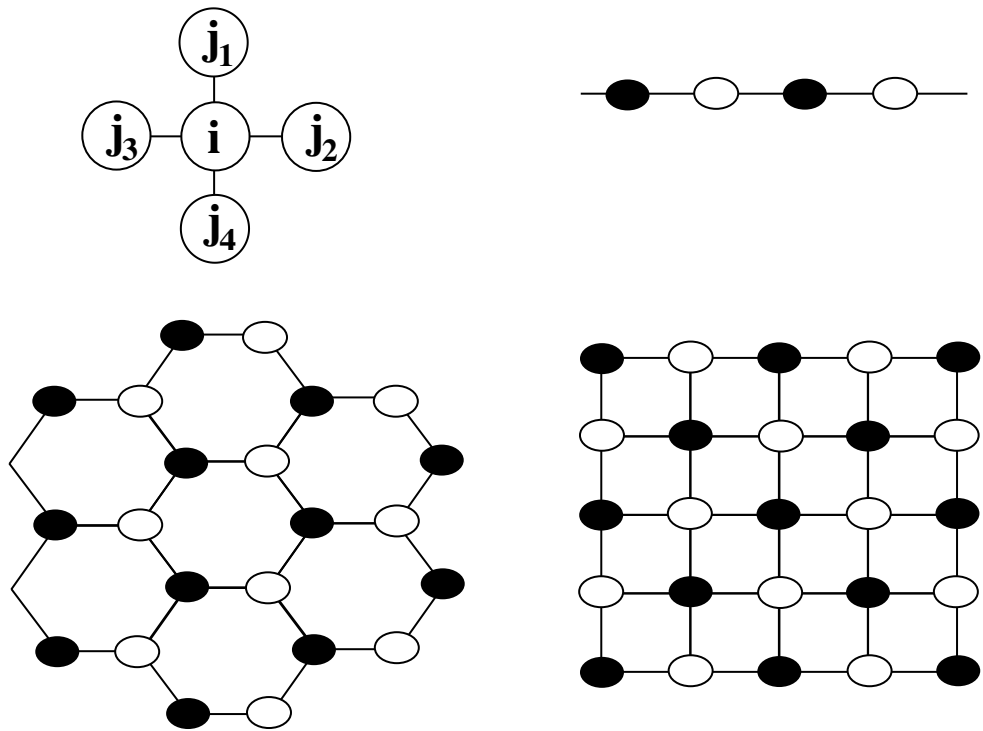}}
\caption{
Interleaved sublattices labeled as filled/open circles
on one dimensional, and two dimensional (square/honeycomb) space.
On a square lattice, $j_{1,2,3,4}$ is the nearest neighbors to site $i$. 
}\label{sublattice}
\end{figure}

\begin{figure}\centerline
{\epsfxsize = 6.in \epsffile{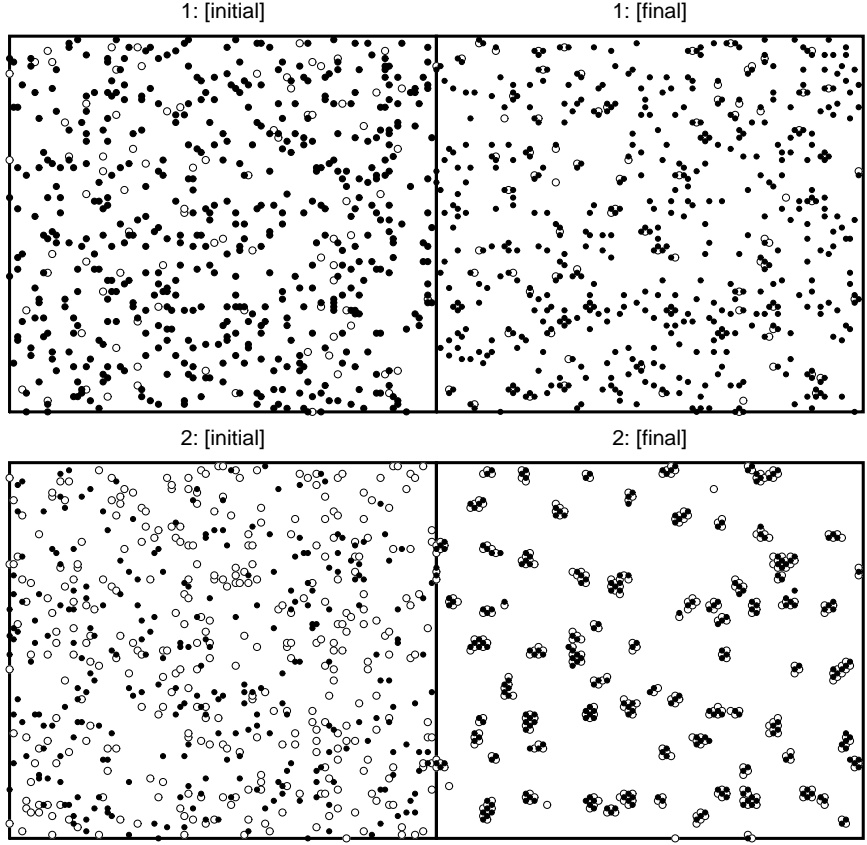}}
\caption{
Monte Carlo simulation with Metropolis scheme. Here we test
the model under a fixed receptor density but different ligand concentrations.
In both upper and lower panels,
the left figure represent initial conditions and the right are results
after $10^8$ Monte Carlo steps. The open, filled circles indicate liganded,
unliganded receptor molecules, respectively.
There is no stable cluster formation in the upper snapshot, 
whereas the clustering is
stable in the lower one. Here we use $g_E=6k_BT$, density of liganded
receptor: upper plane, 0.001, lower plane, 0.03, and density of unliganded
receptor: upper plane, 0.059, lower plane, 0.03.
}\label{monte}
\end{figure}

\begin{figure}\centerline
{\epsfxsize = 6.in \epsffile{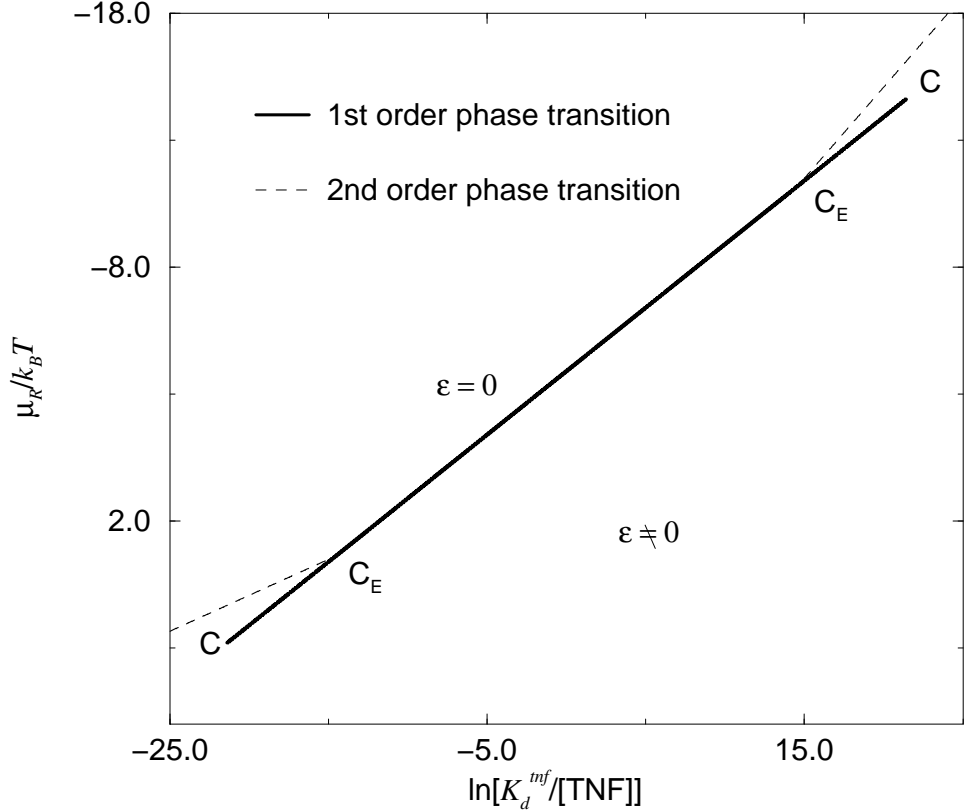}}
\caption{
Numerically computed phase diagram, showing that there is
a pair of second-order lines each of which ends on the first-order
transition curve. (C=critical point; $\rm C_E$=critical endpoint). The phase 
to the lower-right has $\epsilon \neq 0$. Here we used $g_E=6k_BT$ and $D=3$.
To show the symmetry, we plot the ligand concentration in logarithm unit, 
normalized with respect to the dissociation constant $K_d^{tnf}=0.59$ nM.
Here the chemical potential $\mu_R$ is related to receptor density. In 
fig. \protect\ref{phase_separation}, we convert the receptor chemical potential
into the molecular density.}
\label{phase_diagram}
\end{figure}

\begin{figure}\centerline
{\epsfxsize = 6.in \epsffile{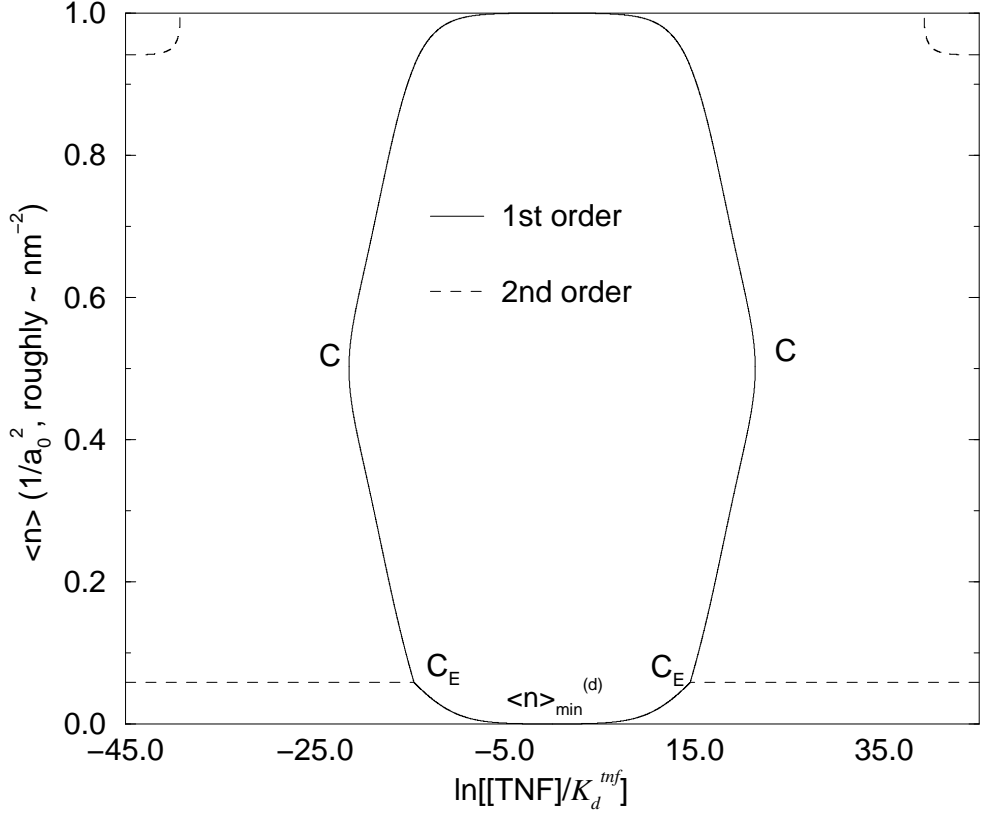}}
\caption{
The phase diagram now shown as a function of
the receptor density $\langle n\rangle$ related to the ligand concentration.
To show the symmetry, we plot the ligand concentration in logarithm unit,
normalized with respect to the dissociation constant $K_d^{tnf}=0.59$ nM.
The region inside the solid lines is the co-existence
region where states of high and low density co-exist. As the ligand
concentration is altered so as to cross one of these lines, the
receptors will spontaneously cluster and thereby allow signaling to
occur. $\langle n\rangle_{min}^{(d)}$ is the minimal receptor density for 
clustering. For this set of parameters, clustering
will occur even for very small overall receptor density. Here $a_0$ is the
length scale for the lattice spacing. For surface receptor molecule such as
TNF-R1, we might take $a_0\approx 1$ nm. 
}\label{phase_separation}
\end{figure}

\begin{figure}\centerline
{\epsfxsize = 6.in \epsffile{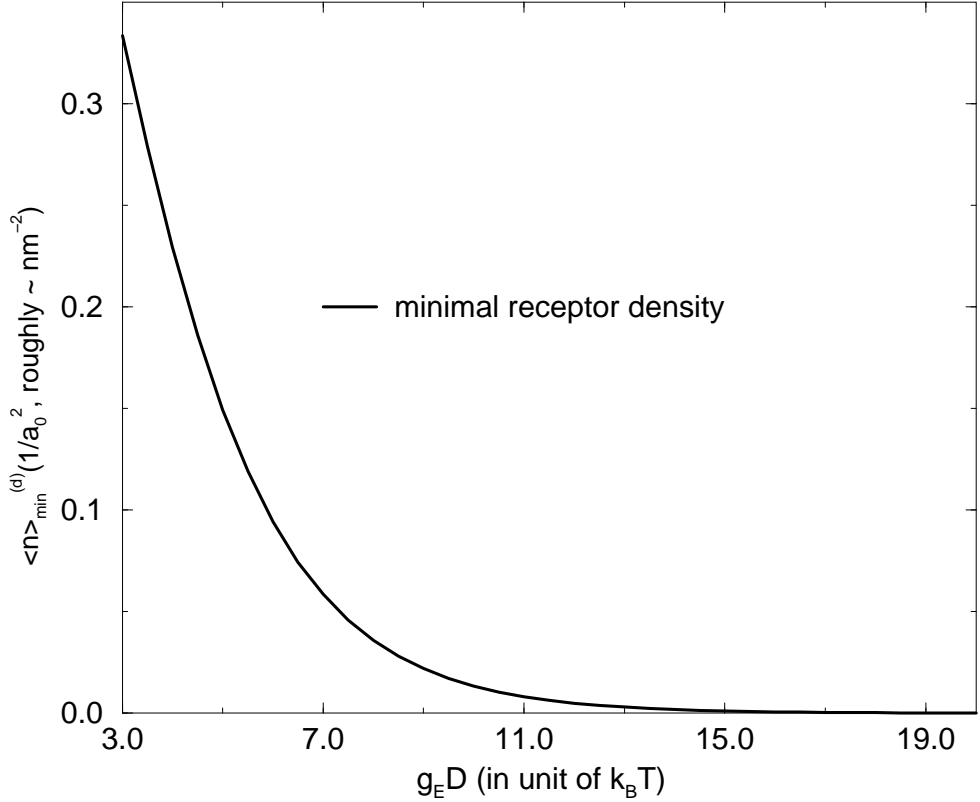}}
\caption{
The variation of $\langle n\rangle_{min}^{(d)}$
as a function of the association energy 
$g_E$. We find that $\langle n\rangle_{min}^{(d)}$ rapidly approaches zero
once $g_ED\ge 15k_BT$. If we assign $D=3\sim 4$, this energy scale 
corresponds to a single hydrogen bond. Here $a_0$ is the
length scale for the lattice spacing. For a surface receptor molecule such as
TNF-R1, we might take $a_0\approx 1$ nm. 
}\label{nmin}
\end{figure}

\end{document}